\documentclass[sigconf,screen]{acmart}
\copyrightyear{2026}
\acmYear{2026}
\setcopyright{cc}
\setcctype{by}
\acmConference[ICCAD '26]{IEEE/ACM International Conference on Computer-Aided Design}{November 08--12, 2026}{San Jose, CA, USA}
\acmBooktitle{IEEE/ACM International Conference on Computer-Aided Design (ICCAD '26), November 08--12, 2026, San Jose, CA, USA}
\acmDOI{10.1145/3831252.3834067}
\acmISBN{979-8-4007-2873-0/2026/11}

\usepackage{enumitem}
\setlist{leftmargin=5.08mm}

\usepackage{lipsum}
\usepackage{graphicx}
\usepackage{amsmath}
\usepackage{footnote}
\usepackage{mathtools}
\usepackage{mathrsfs}     
\usepackage{comment}
\usepackage{soul}                                          
\usepackage[subrefformat=parens,labelformat=parens]{subfig}
\usepackage{bm,bbm}
\usepackage{multirow}
\usepackage{threeparttable,booktabs}
\usepackage{blkarray}
\usepackage{tikz}
\usetikzlibrary{positioning,calc,fit,decorations.pathmorphing,shapes.geometric,shapes.gates.logic.US,calc}
\usetikzlibrary{arrows,arrows.meta,decorations.markings,shapes,shapes.arrows}
\usetikzlibrary{patterns,snakes}
\usetikzlibrary{decorations,decorations.pathreplacing}
\usetikzlibrary{backgrounds}
\usepackage{tikz-3dplot}
\usepackage{balance}
\usepackage{courier}                                       
\usepackage{hyperref}
\usepackage[nameinlink]{cleveref}                                   
\usepackage[mathcal]{eucal}
\usepackage[]{algpseudocode}                               
\algrenewcommand\textproc{\texttt}
\makeatletter\let\float@addtolists\relax\makeatother
\usepackage{algorithm}
\usepackage{filecontents}                                  
\usepackage{pgfplots}
\usepackage{pgfplotstable}
\usepackage{pgf-pie}
\usepgfplotslibrary{groupplots}
\pgfplotsset{compat=newest}
\usepackage[figuresright]{rotating}
\usepackage{xcolor,colortbl}                               
\usepackage{adjustbox}                                     
\usepackage{makecell}
\usepackage{siunitx}
\usepackage{color}
\usepackage{diagbox}

\newcommand{\minisection}[1]{\vspace{.06in}\noindent{\textbf{#1}}.}

\theoremstyle{plain}

\theoremstyle{definition}

\algrenewcommand\textproc{\texttt}

\usepackage[skip=1pt]{caption}            
\definecolor{CUHKorange}{RGB}{244,106,18} 
\definecolor{CUHKblue}{RGB}{0,111,190}    
\definecolor{CUHKgreen}{RGB}{0,127,128}   
\definecolor{CUHKred}{RGB}{228,46,36}     
\definecolor{CUHKyellow}{RGB}{198,148,34} 
\definecolor{CUHKdark}{RGB}{114,44,114}   
\definecolor{CUHKmiddle}{RGB}{144,44,144} 

\usepackage{tcolorbox}
\tcbuselibrary{skins,breakable}
    {\endtcolorbox}
    {\endtcolorbox}

\graphicspath{{./figs/}{../}}

\makeatletter
\newcommand{\manualauthornote}[1]{%
  \g@addto@macro\@authornotes{%
    \stepcounter{footnote}\footnotetext{#1}}}
\makeatother

\begin{document}
\date{}

\title{
    SGLayout: Schema-Guided Closed-Loop Refinement for LLM-Driven Parameterized Layout Generation
}

\author{\Large
Siyang Chen\textsuperscript{1,2},
Yuan Pu\textsuperscript{3},
Yuhan Qin\textsuperscript{3},
Zhaotan Lin\textsuperscript{3},
Siyuan Liang\textsuperscript{3},
Zhuolun He\textsuperscript{3},
Lining Zhang\textsuperscript{1*},
Bei Yu\textsuperscript{3}
}
\affiliation{
  \vspace{1pt}
  \institution{\large
  \textsuperscript{1}School of Electronic and Computer Engineering, Peking University, Shenzhen, China\\[1pt]
  \textsuperscript{2}ChatEDA Tech, Shenzhen, China\\[1pt]
  \textsuperscript{3}Department of Computer Science and Engineering, The Chinese University of Hong Kong, Hong Kong SAR}
  \city{}
  \country{}
}
\manualauthornote{Corresponding author.}
\renewcommand{\shortauthors}{Siyang Chen et al.}

\settopmatter{printfolios=false}

\begin{abstract}

Large Language Models (LLMs) have shown strong potential in parameterized layout generation via tool-augmented code synthesis. However, LLMs lack spatial awareness, making one-shot generation unreliable. While layout quality can be evaluated with established techniques, a critical gap remains: geometric-domain evaluation outcomes cannot be directly translated into actionable code-level fixes, trapping iterative refinement in blind refactoring. We propose SGLayout, a self-refining framework that closes this gap via a schema-guided mechanism. Upon code execution, a structured schema captures the correspondence between each code-level instance and its geometric attributes, establishing a bidirectional link between code space and geometry space. Emulating engineer review practices, SGLayout evaluates layouts along two complementary dimensions: intent compliance through quantitative structural checking and qualitative vision-language assessment, and rule compliance through DRC. All identified violations are then back-projected through the schema onto precise code targets, enabling surgical repair. This closed-loop process iterates until layout convergence. On a curated benchmark of 112 parameterized layout generation tasks using GDSFactory, SGLayout achieves a 76.79\% pass rate, compared with 50.89\% for the strongest baseline.
 
\end{abstract}

\maketitle
\section{Introduction}
\label{sec:intro}

Large language models (LLMs) have recently demonstrated strong capabilities in program synthesis and tool-augmented reasoning, motivating growing interest in their application to electronic design automation (EDA). Among emerging directions, parameterized layout generation is especially promising. In this paradigm, physical design is represented as executable programs in programmatic layout environments such as GDSFactory \cite{gdsfactory}, where code instantiates device primitives and specifies spatial relations, enabling reusable and automated design in domains like silicon photonics, analog circuits and mixed-signal design, where both geometric regularity and parameter flexibility are essential.

\begin{figure}[!t]
    \centering
    \subfloat[\label{fig:intro-oneshot}]{
        \includegraphics[width=.92\columnwidth,trim=20 10 20 10,clip]{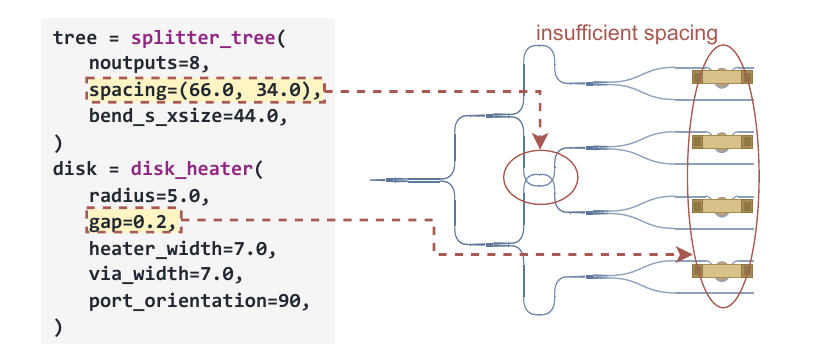}
    }\\[0.15em]
    \subfloat[\label{fig:intro-blind}]{
        \includegraphics[width=.92\columnwidth,trim=20 10 20 10,clip]{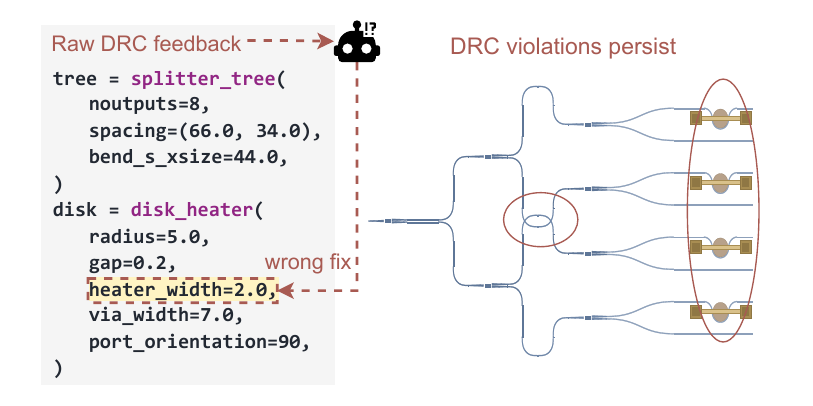}
    }
    \caption{Examples of failures in LLM-driven parameterized layout generation due to the limited spatial awareness. (a) One-shot generation. (b) Raw DRC feedback leads to blind refactoring.}
    \label{fig:intro-motivation}
\end{figure}

Parameterized layout generation formulates physical design as a code generation task, allowing LLMs to leverage their program synthesis capabilities to directly translate natural-language specifications into executable layout programs. However, because layout is generated in code space while correctness is judged in geometry space, LLMs struggle to anticipate spatial consequences. Therefore, the initial one-shot generated layout often fails, presenting spatial errors that violate design intent or physical constraints. \Cref{fig:intro-oneshot} shows a one-shot LLM-generated 8-way splitter tree with interleaved disk-heater and reference branches; the LLM selects insufficient spacing, and the disk-heater and its adjacent waveguide remain too close due to insufficient gap, causing minimum-spacing violations.

Recently, a growing body of prior work has explored LLMs for EDA, including field-level overviews \cite{llmedasurvey2025,he2025futuremirage}, HDL generation and benchmarking \cite{chipnemo2023,chipgpt2023,verilogeval2023,rtllm2024,verigen2023,betterv2024}, knowledge-grounded LLM systems, RAG-assisted documentation QA, and autonomous design agents \cite{infusingeda2026,customragedaiccad2024,mmgrade2025,maeda2026,wu2024chateda,rtlfixer2023,hdldebugger2025,jarvis2025}. In physical design, prior efforts have explored layout automation, optimization, and verification \cite{scheible2015,magical2019,align2020,gpubuffer2025,blade2026,llmstdcell2024,drccoder2024,layoutcopilot2025,you2024primitive}. Beyond these tasks, recent advances have also extended LLMs toward natural-language-driven layout synthesis. In particular, GLayout \cite{glayout2024} translates analog layout into a compact, technology-generic text representation and uses LLMs to support analog layout automation. PhIDO \cite{phido2025} targets photonic integrated circuit (PIC) layout generation via a staged pipeline in which LLM agents translate user intent into a structured DSL and further instantiate it within a library-supported design space composed of curated templates and PDK components. More importantly, PhIDO itself shows that reliability degrades sharply on complex tasks: on Level 4 problems, the best-performing model reaches only 8.9\% pass@1. This indicates that one-shot synthesis, as illustrated in \Cref{fig:intro-oneshot}, remains unreliable for realistic complex layouts, motivating the need for a closed-loop refinement framework. However, the refinement loop still faces a fundamental ungrounded feedback problem: due to the representation gap between code space and geometry space, a reported spatial violation rarely reveals which code snippet or parameter is actually responsible, forcing the LLM into blind refactoring rather than targeted repair. As demonstrated in \Cref{fig:intro-blind}, given raw DRC feedback, the LLM modifies a width parameter that appears relevant to the violation, rather than identifying and fixing its true root cause.

Therefore, we propose SGLayout, a framework for schema-guided closed-loop refinement in LLM-driven parameterized layout generation. The key idea is to use an execution-derived structured schema as the alignment layer between program structures and geometric realizations. After each generated layout program is executed, the system extracts a schema that records the correspondence between code-level instances and their geometric attributes, establishing a bidirectional link between code space and geometry space. Built on this representation, the framework evaluates generated layouts along two complementary dimensions: intent compliance, assessed through quantitative API-based structural checking and qualitative vision-language analysis, and rule compliance, assessed through DRC. More importantly, the schema serves as the common bridge that maps these heterogeneous findings into actionable code targets. The framework localizes feedback findings through the schema into explicit repair targets, enabling grounded and targeted refinement until the layout converges. By explicitly linking evaluation outcomes back to their source code instances, our schema transforms ungrounded guessing into surgical repair, effectively breaking the cycle of blind refactoring illustrated in \Cref{fig:intro-motivation}.

To systematically evaluate the capabilities of SGLayout and promote the development of automated parameterized layout generation, we construct a comprehensive benchmark comprising 112 high-quality and real-scenario parameterized layout generation tasks implemented in GDSFactory. These tasks cover diverse geometries and require the generated code to strictly satisfy both design intent and physical rules. The benchmark is open-source at \url{https://github.com/Zeplll/SGLayout-benchmark}.

We summarize our key contributions as follows: 
\begin{itemize}
    \item We propose SGLayout, a schema-guided closed-loop refinement framework for LLM-driven parameterized layout generation. To address the limited spatial awareness of LLMs, SGLayout leverages multi-dimensional evaluators for self-assessment and a schema-guided mechanism for self-refinement, enabling layouts to satisfy both design intent and physical constraints.
    \item We introduce an execution-derived schema as the core cross-modal alignment mechanism. By bridging the fundamental gap between code space and geometry space, the schema maps heterogeneous evaluation findings to precise code targets for surgical repair.
    \item We construct a comprehensive benchmark comprising 112 high-quality and real-scenario parameterized layout generation tasks.
    \item We conduct extensive experiments showing that SGLayout achieves 76.79\% pass rate on our curated benchmark, significantly outperforming the strongest baseline.
\end{itemize}

\section{Preliminaries}
\label{sec:prelim}
\subsection{Parameterized Layout}
In photonic, analog, and mixed-signal physical design, layouts are increasingly developed through programmable generators rather than fully manual drawing. This design style is particularly suitable for domains that require both structural regularity and parameter flexibility, and is commonly supported by programmatic layout environments such as GDSFactory \cite{gdsfactory}. The central building block in this paradigm is the Parameterized Cell (PCell)---a reusable generator that produces device geometries from input variables \cite{farsaei2023pcells}. In practice, however, a complete parameterized layout is not limited to isolated PCell instances; it also involves hierarchical component composition, relative placement, routing and interconnect construction, and custom geometric primitives created through layout APIs.

Given a natural-language specification $q$, a layout program $C$ instantiates components, places references, constructs paths and polygons, and exports a physical layout. Such a parametric and component-centric design style is also consistent with established photonic design frameworks such as IPKISS \cite{bogaerts2012ipkiss}. We refer to this symbolic and hierarchical program representation as the code space $\mathcal{C}$. Executing $C$ yields a concrete layout $G=\mathrm{Exec}(C)$, represented by geometry-level entities such as instances, polygons, ports, coordinates, layers, and rendered views; we refer to this realized representation as the geometry space $\mathcal{G}$. Parameterized layout generation is therefore performed in code space, while correctness is assessed in geometry space through structural checking, visual inspection, and design-rule checking (DRC). This separation gives rise to a representation gap between program semantics and geometric realization.

\subsection{Problem Formulation}
Given a natural-language specification $q$, the objective is to synthesize code $C \in \mathcal{C}$, and execute it into geometric layout $G = \text{Exec}(C) \in \mathcal{G}$. The generated layout is then evaluated by a set of multi-dimensional evaluators, producing evaluation findings $F$ that may include structural checking failures, visual findings, and DRC violations:
$$F = \text{Eval}(G).$$

Let $T$ denote the localized code-level faults responsible for these evaluation findings $F$, serving as actionable repair targets. The objective is to iteratively refine the generated program in code space:
$$C_{t+1} = \text{Repair}(C_t, T_t).$$

This requires a backward mapping function $\Phi$ which maps the geometry-space evaluation findings $F$ back to their corresponding repair targets $T$ within the source code $C$:
$$\Phi: (F,C) \rightarrow T.$$

Without $\Phi$, the representation gap between code space and geometry space cannot be closed, leaving iterative refinement to operate blindly.
\section{Schema-Guided Mechanism}
\label{sec:method1}

\subsection{Schema Construction}
\label{sec:schema-construction}
To bridge the representation gap between code space and geometry space, SGLayout constructs an execution-derived schema which captures the structural attributes of layout entities and their associated code snippets through a dual-source acquisition mechanism during the forward execution flow $G = \text{Exec}(C)$, as shown in the forward portion of \Cref{fig:schema}. 

The construction process is realized by first performing semantic metadata extraction, where the system invokes parameterized layout APIs after code generation to extract intent-level metadata, including hierarchical cell nesting, transformation matrices, port-level connectivity, configuration parameters of components, geometric properties of polygons and paths, among others. Simultaneously, the system performs instruction-level runtime interception via dynamic hooks on every geometric instantiation call. This mechanism captures each generated spatial primitive such as a polygon or an instance and directly records its originating code snippet into the corresponding element's entry within the schema. Through this provenance record, the schema acts as a bidirectional lookup table that maps each generated component to its source code snippet and configuration parameters. This explicit linkage enables subsequent error back-projection from geometry space to code-level repair targets.

\subsection{Schema-Guided Error Back-Projection}
\label{sec:back-projection}
\begin{figure}[!t]
    \centering
    \includegraphics[width=\columnwidth,trim=20 15 20 12,clip]{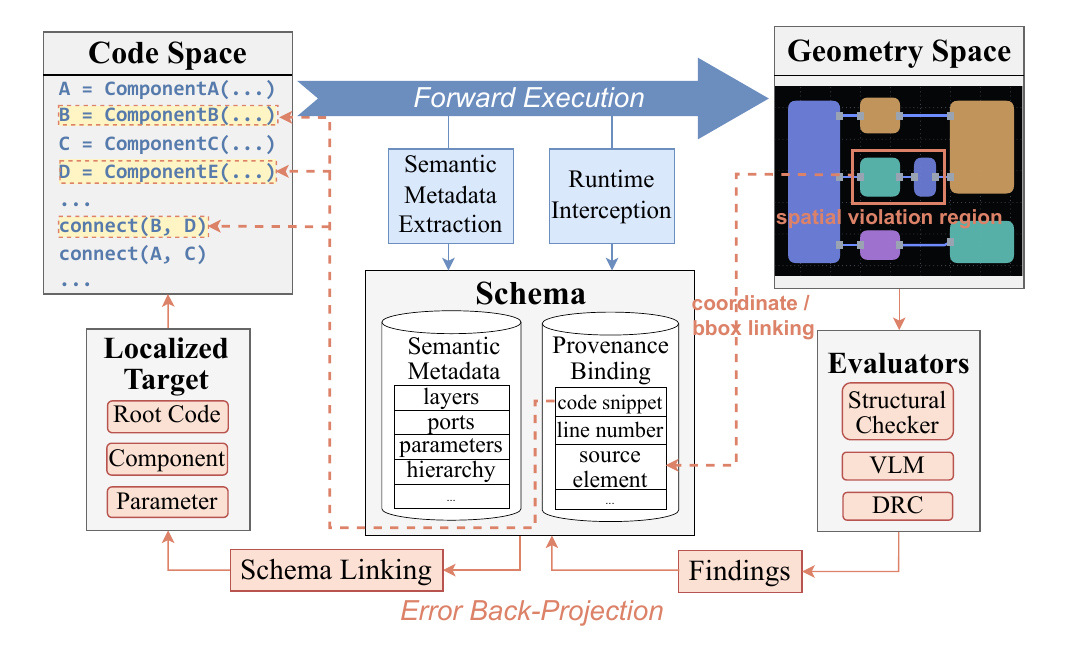}
    \caption{Illustration of the schema-guided mechanism.}
    \label{fig:schema}
\end{figure}

Although the construction of the schema successfully aggregates the semantic and structural information of the code space $\mathcal{C}$, evaluation findings are still produced in the geometry space $\mathcal{G}$. As shown in the backward portion of \Cref{fig:schema}, SGLayout closes this gap through a schema-guided error back-projection mechanism. This mechanism serves as the concrete realization of the backward mapping function $\Phi: F \times C \rightarrow T$ defined in our problem formulation, converting heterogeneous, geometry-centric evaluation findings $F$ into actionable, localized code-level targets $T$.

Given a normalized finding from the evaluation layer, SGLayout first identifies the responsible schema entity from the finding evidence. For structural findings, this mapping is direct because the API-based structural checker queries the schema itself and any failed check inherently identifies the violating schema entity. Furthermore, because the checker is generated from intent decomposed contracts, each failure also explicitly corresponds to a specific contract unit and its associated code block. For spatial findings produced by the VLM or DRC, localization proceeds through the geometric evidence associated with the violations, including reported coordinates, bounding boxes, and overlap regions. SGLayout computes the spatial overlap between this violation-level geometric evidence and the bounding boxes recorded for schema entities, thereby linking the violation to the corresponding instance, path, or polygon in the schema. Once a finding is mapped to the relevant schema entities, SGLayout traces it back to the corresponding code anchor through the provenance binding information stored in the schema. 

In this way, schema-guided error back-projection transforms abstract, ungrounded evaluation feedback into precise, code-level targets for subsequent repair. By mapping spatial failures back to their error-inducing code regions, the $\Phi$ function effectively breaks the cycle of blind refactoring and ensures that the refinement loop is driven by surgical edits.

\section{Proposed Framework}
\label{sec:method2}

\begin{figure*}[t]
\centering
\includegraphics[width=.92\textwidth,trim=20 5 20 135,clip]{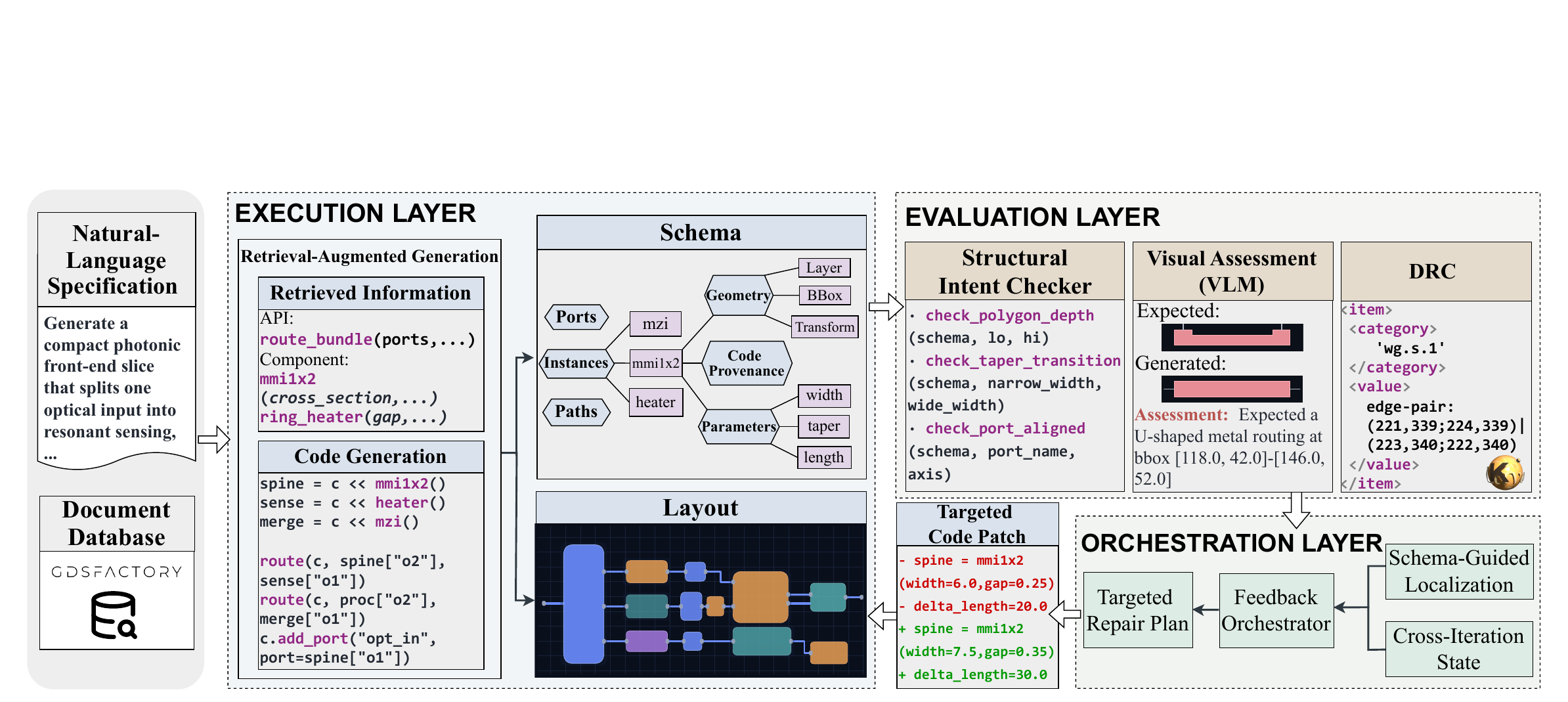}
\caption{Overview of the SGLayout framework.}
\label{fig:overview}
\end{figure*}

\subsection{Overview}
While the schema-guided mechanism provides the theoretical foundation for cross-modal alignment, translating it into a practical refinement framework requires a robust system architecture. Naive LLM-driven repair loops rely on untargeted trial-and-error, leading to unstable refinement trajectories and repeated failure to resolve geometry-space violations.

To overcome this, we propose SGLayout, a schema-guided closed-loop refinement framework for LLM-driven parameterized layout generation, as illustrated in \Cref{fig:overview}. Instead of treating repair as a stateless regeneration process, the framework organizes refinement around the schema, which serves as the shared structural backbone linking program structure with geometric realization across iterations. Given a natural-language specification, the execution layer first performs retrieval-augmented generation to produce an executable layout program, materialize its physical layout, and extract the corresponding schema. The evaluation layer then examines the generated result through multi-dimensional evaluators, exposing violations in both intent compliance and rule compliance and converting them into a normalized set of repair findings. Based on these findings, the orchestration layer localizes each failure to concrete code targets through the schema, reasons over cross-iteration state signals such as schema diff and constraint ledger, and synthesizes a targeted repair plan. Finally, guided by the plan, SGLayout performs closed-loop refinement by applying constrained local patches to the previous program. In this way, SGLayout transforms raw geometry-space feedback into grounded code-level repair, establishing a principled closed loop for iterative parameterized layout generation.

\subsection{Execution Layer}
The execution layer begins with a retrieval-augmented generation stage specialized for GDSFactory programming. SGLayout retrieves from a unified knowledge base of API documents, component descriptions, and curated notebook-style examples, all preprocessed into fine-grained document chunks. Given a user query, the retriever first performs dense recall, then applies type-aware filtering and reranking to retain a compact set of task-compatible references. The final retrieved context therefore serves as concise implementation evidence, rather than a naive concatenation of nearest-neighbor chunks.

The code generation agent consumes this retrieved context together with the natural-language specification. In the initial round, it uses the retrieved chunks as implementation evidence for selecting valid constructors, parameters, and composition patterns. The generated program is then executed to materialize the physical layout, while the corresponding schema introduced in \Cref{sec:schema-construction} is extracted through semantic metadata extraction and runtime interception during the forward execution to provide the structural backbone for downstream evaluation and refinement. Retrieval is further updated across refinement rounds: when evaluation exposes missing API knowledge or misuse patterns, SGLayout performs feedback-augmented supplementary retrieval to supply additional context for the next repair step. Consequently, SGLayout ensures retrieval-augmented generation, effectively preventing hallucination.

\subsection{Evaluation Layer}
To ensure that the generated layout satisfies both design intent and physical validity, SGLayout adopts a multi-dimensional evaluation layer. Relying solely on a single evaluation dimension is insufficient; a layout might pass rule checks but fail to implement the required circuit topology, or conversely, match the intent but violate minimum spacing rules. Therefore, our framework evaluates the layout across two complementary dimensions—intent compliance and rule compliance—through three specialized evaluators:

\minisection{API-based Structural Intent Checker}
The API-based structural intent checker is fully decoupled from the generated code, operating solely based on the standardized schema format and the natural-language query. It adopts a two-layer design. At the lower layer, human engineers define a reusable library of schema-native checking templates, implemented as API-level checking functions over the execution-derived schema. The library contains 74 template functions grouped into eight semantic categories, covering structural constraints such as layer usage, instance and port counts, geometric relations, routing properties, hierarchical reuse, and parameter overrides. At the upper layer, SGLayout uses LLM agents to automatically instantiate a structural checker for each query. An intent-decomposition agent translates the user specification into independently verifiable contracts. These contracts are then mapped to relevant templates and composed into an executable evaluator. Custom logic is introduced only when template-based specification is insufficient. This design removes the need for manual per-query test writing while preserving the precision and interpretability of API-based structural verification, thereby improving both scalability and consistency.

\minisection{Vision-Language Assessment}
We employ the VLM as a complementary evaluator for macro-level visual intent. While the structural checker can verify many symbolic constraints, it is less effective at capturing holistic visual plausibility, such as gross asymmetry, unnatural routing detours, or malformed global shapes. In the absence of a schematic representation suitable for traditional graph-based algorithms, the VLM provides a heuristic visual assessment of such macro-level intent violations. To improve both assessment quality and spatial grounding, we provide the VLM with two complementary renderings of each layout: a raw view that preserves the geometric appearance of the design, and an annotated view that adds coordinates and semantic markers such as ports and instances. The raw view supports holistic perception, whereas the annotated view helps the model return more grounded spatial references. Importantly, the scope of the VLM evaluator is intentionally limited to qualitative visual assessment, rather than exact numerical or coordinate-level verification.

\minisection{Design Rule Checking}
A standard Design Rule Checking (DRC) engine is invoked to validate whether the layout is physically legal under fabrication rules. The generated DRC report is further parsed into a structured form, from which rule types, violation counts, and violation regions are extracted, and these regions are linked back to schema entities through spatial overlap, enabling subsequent schema-guided back-projection and repair.

\medskip
\noindent
To prevent the refinement loop from being overwhelmed by excessive detailed feedback, SGLayout organizes these evaluators in a coarse-to-fine hierarchical sequence, prioritizing structural intent checking, followed by visual assessment, and finally fine-grained DRC. If a layout fails at a coarser evaluation tier, the process halts early and directly forms the corresponding feedback for repair, rather than proceeding to lower-level checks whose results may be secondary symptoms. Since the resulting evaluator outputs are inherently heterogeneous, ranging from failed programmatic assertions to spatial violation regions and raw reports, the evaluation layer further converts them into normalized findings ($F$) that preserve the structural or spatial evidence required by downstream repair planning. The reliability of these automatic evaluators is assessed separately in \Cref{sec:evaluator-reliability}.

\subsection{Orchestration Layer}
\begin{figure}[t]
\centering
\includegraphics[width=\columnwidth,trim=20 5 20 5,clip]{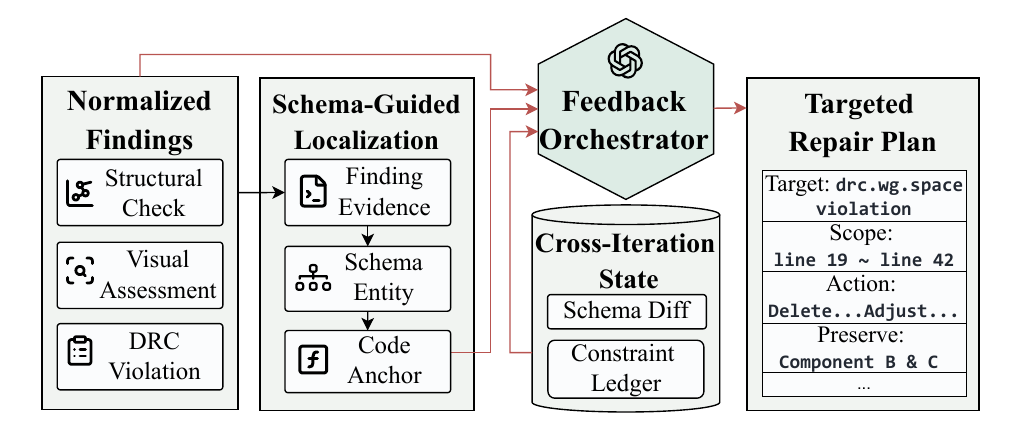}
\caption{Orchestration Layer: From Normalized Findings to Targeted Repair Plan.}
\label{fig:orchestrator}
\end{figure}

The orchestration layer transforms normalized findings ($F$) into a targeted repair plan, as illustrated in \Cref{fig:orchestrator}. Starting from the normalized findings produced by the evaluation layer, SGLayout first converts each finding into a localized repair target through schema-guided back-projection as described in \Cref{sec:back-projection}: the finding evidence is linked to the responsible schema entity, and the corresponding code anchor is then recovered through provenance binding. The orchestration layer further augments these localized signals with cross-iteration state, and uses a LLM-based feedback orchestrator to synthesize a compact repair plan that specifies what to edit, where to edit, and how to edit. Rather than passing raw feedback directly to the code agent, the orchestration layer gives SGLayout the ability to reason over the source of the feedback, identify the root cause behind observed violations, and derive repair actions that operate on the true error-inducing code region.

\minisection{Cross-Iteration State Modeling}
A key property of SGLayout is that refinement is stateful across iterations rather than a sequence of independent regeneration attempts. To support repair planning over time, the framework maintains two forms of cross-iteration state: a schema diff that captures structural progress signals, and a constraint ledger that records cross-round failure memory.

To expose whether the latest repair has produced a meaningful structural effect, the framework compares the current execution schema $S_t$ with that of the previous iteration $S_{t-1}$ and computes an iteration-level structural delta:
$$\Delta S_t = S_t - S_{t-1}$$
In our implementation, this schema diff tracks a compact set of schema-level observables, including instance bounding boxes and the total numbers of instances, ports, and paths. These signals are sufficient to reveal whether a repair has actually changed the instantiated structure, adjusted placements, restored missing connectivity artifacts, or inadvertently removed required routing content. $\Delta S_t$ provides the feedback orchestrator with a direct progress signal for detecting the trend of modifications and ineffective edits.

SGLayout further maintains a constraint ledger as a cross-round memory of normalized findings. The ledger assigns each finding a stable constraint identifier and records its historical status, including first occurrence, last pass, last failure, and, when unresolved, the trend of its error values over time. This allows the feedback orchestrator to reason over the trajectory of constraint satisfaction rather than reacting only to the current round’s failures. As a result, the framework can better avoid reintroduction of previously resolved violations and oscillation among tightly coupled constraints.

\minisection{Feedback Orchestration}
Even after normalization and schema-guided localization, the resulting feedback may still be verbose and uneven in causal importance: some findings reflect root causes, whereas others are only downstream symptoms. Directly feeding all localized findings back to code generation therefore often results in shallow and inefficient repairs, as the model may react to downstream symptoms without identifying the true root cause. To address this, SGLayout introduces a feedback orchestrator, implemented as an LLM agent, that converts localized findings into a compact and structured repair plan, making SGLayout's closed loop practically effective.

Given the normalized findings, code anchors, schema diff, and constraint ledger, the orchestrator deduplicates redundant errors, prioritizes findings by severity and persistence, and abstracts each selected finding into a structured repair contract that defines the edit scope, repair action, and preservation constraints, among others. This design constrains the model to operate on the smallest causal block, rather than drifting into global rewrites. By reasoning jointly over current evidence and cross-iteration state, it can also detect persistent violations and shift the repair focus from local geometric tweaks to the underlying root cause.

\medskip
\noindent
With the repair plan produced by the orchestration layer, SGLayout performs the next refinement round by applying constrained local patches to the previously generated executable program, rather than regenerating the entire layout from scratch. In this way, SGLayout forms a complete closed loop over generation, execution, evaluation, localization, orchestration, and repair, turning iterative refinement from blind refactoring into a targeted and systematic repair process. The loop terminates when all evaluation stages pass, or when the maximum number of retries is reached.
\section{Experimental Results}
\label{sec:experimental-results}

\subsection{Evaluation Benchmark}
We evaluate SGLayout on a curated benchmark of 112 specification-driven parameterized layout generation tasks implemented in GDSFactory. Rather than assuming a unique hidden reference layout for each task, the benchmark pairs each natural-language query with an executable evaluation protocol, reflecting the open-ended nature of parameterized layout generation where multiple implementations may satisfy the same specification. 
The benchmark includes 112 natural-language queries, one structural intent checker for each query, and a reusable library of checker template functions. Although the reliability results in \Cref{sec:evaluator-reliability} support automatic structural checking, the benchmark still releases manually reviewed and corrected per-query structural intent checkers, produced by the structural-checking agent pipeline, to ensure stable and reproducible evaluation. 
We further characterize the benchmark along two orthogonal axes: semantic group and complexity level. The semantic groups are \emph{Geometric Shape Synthesis}, \emph{Connectivity and Routing Construction}, and \emph{Hierarchical and Parametric Composition}, covering custom polygon construction and manufacturability cleanup, port-aware placement and routing, and hierarchical and parameterized composition, respectively. Orthogonally, the complexity levels range from local single-device tasks to multi-instance composition tasks and further to hierarchy-heavy or globally coupled tasks with multiple interacting intents. \Cref{tab:benchmark} summarizes the distribution across these categories. The benchmark is open-source at \url{https://github.com/Zeplll/SGLayout-benchmark}.

\begin{table}[t]
\centering
\caption{Distribution of the benchmark.}
\label{tab:benchmark}
\footnotesize
\begin{tabular}{l c c c c}
\toprule
Semantic Group & Total & Level 1 & Level 2 & Level 3 \\
\midrule
Geometric Shape Synthesis & 26 & 9 & 14 & 3 \\
Connectivity and Routing Construction & 56 & 13 & 28 & 15 \\
Hierarchical and Parametric Composition & 30 & 4 & 4 & 22 \\
\midrule
Total & 112 & 26 & 46 & 40 \\
\bottomrule
\end{tabular}
\end{table}

\subsection{Experimental Settings}
In our primary evaluation, all methods share the same execution environment, the same end-to-end pass criterion and the same maximum budget of five refinement rounds. To ensure the fairness and stability of evaluators, all compared methods are evaluated with the same released query-specific structural intent checkers, the same VLM agent based on \texttt{Qwen-VL-Max}, and the same DRC setup based on GDSFactory Generic PDK \cite{gdsfactorygenericpdk} throughout the benchmark. A sample is counted as solved only if the generated code passes sandbox execution, structural intent checking, VLM, and DRC review. If any stage fails, the process halts and enters the next refinement round, continuing until all stages pass or the retry budget is exhausted.

The full SGLayout system uses \texttt{Qwen3-Coder-Plus} as the code-generation backbone. For comparison, we evaluate two baseline families, namely the naive baseline and PhIDO \cite{phido2025}, each instantiated with \texttt{Qwen3-Coder-Plus}, \texttt{GPT-5.1}, and \texttt{DeepSeek- \allowbreak V3.2} under the same evaluator stack and retry budget. To isolate the effect of the repair stack, SGLayout and the naive baseline share the same execution layer and evaluation layer; the difference is that only SGLayout enables the orchestration layer whereas the naive baseline directly feeds raw feedback back to the generator. PhIDO is included as a domain-matched baseline because it is likewise an LLM-driven photonic layout generation pipeline based on library-grounded component retrieval and composition. Since the PhIDO pipeline depends on library coverage, we manually extend the released design library with missing benchmark-specific components and geometric helpers to keep it executable on our benchmark.

In \Cref{tab:main-results}, PhIDO’s Pass@3 and Final are based on three and five independent generation attempts rather than iterative repair: a query is counted as solved if any of the first three or five repeated PhIDO runs succeeds, whereas for SGLayout and the naive baseline they denote convergence within three and five repair rounds of a single run. These repeated PhIDO runs are included to mitigate the effect of LLM generation stochasticity, rather than to perform stateful refinement.

\subsection{Results}

\begin{table*}[t]
\centering
\caption{Performance of SGLayout, naive baselines, and PhIDO on the 112-task benchmark. Final denotes Pass@5, corresponding to up to 5 repair rounds for SGLayout and the naive baselines, and 5 independent runs for PhIDO.}
\label{tab:main-results}
\footnotesize
\setlength{\tabcolsep}{2.6pt}
\begin{tabular}{>{\centering\arraybackslash}p{2.05cm} >{\centering\arraybackslash}p{2.55cm} ccc ccc ccc ccc}
\toprule
\multicolumn{1}{c}{\multirow{2}{*}{Method}} & \multicolumn{1}{c}{\multirow{2}{*}{Model}} & \multicolumn{3}{c}{\shortstack{Level 1\\Pass Rate (\%)}} & \multicolumn{3}{c}{\shortstack{Level 2\\Pass Rate (\%)}} & \multicolumn{3}{c}{\shortstack{Level 3\\Pass Rate (\%)}} & \multicolumn{3}{c}{\shortstack{Overall\\Pass Rate (\%)}} \\
\cmidrule(lr){3-5}\cmidrule(lr){6-8}\cmidrule(lr){9-11}\cmidrule(lr){12-14}
 &  & Pass@1 & Pass@3 & Final & Pass@1 & Pass@3 & Final & Pass@1 & Pass@3 & Final & Pass@1 & Pass@3 & Final \\
\midrule
Naive & Qwen3-Coder-Plus & 73.08 & 92.31 & 96.15 & 23.91 & 34.78 & 45.65 & 7.50 & 25.00 & 27.50 & 29.46 & 44.64 & 50.89 \\
Naive & DeepSeek-V3.2 & 65.38 & 76.92 & 80.77 & 28.26 & 39.13 & 45.65 & 12.50 & 22.50 & 25.00 & 31.25 & 41.96 & 46.43 \\
Naive & GPT-5.1 & 53.85 & 57.69 & 65.38 & 23.91 & 32.61 & 34.78 & 17.50 & 25.00 & 25.00 & 28.57 & 35.71 & 38.39 \\
PhIDO & Qwen3-Coder-Plus & 42.31 & 61.54 & 65.38 & 36.96 & 41.30 & 41.30 & 17.50 & 25.00 & 27.50 & 31.25 & 40.18 & 41.96 \\
PhIDO & DeepSeek-V3.2 & 34.62 & 46.15 & 46.15 & 30.43 & 36.96 & 39.13 & 10.00 & 20.00 & 20.00 & 24.11 & 33.04 & 33.93 \\
PhIDO & GPT-5.1 & 50.00 & 53.85 & 61.54 & 21.74 & 30.43 & 30.43 & 7.50 & 12.50 & 17.50 & 23.21 & 29.46 & 33.04 \\
\midrule
SGLayout (ours) & Qwen3-Coder-Plus & 61.54 & 88.46 & \textbf{100.00} & 34.78 & 78.26 & \textbf{93.48} & 10.00 & 35.00 & \textbf{42.50} & 32.14 & 65.18 & \textbf{76.79} \\
\bottomrule
\end{tabular}
\end{table*}

\begin{figure}[t]
\centering
\includegraphics[width=\columnwidth]{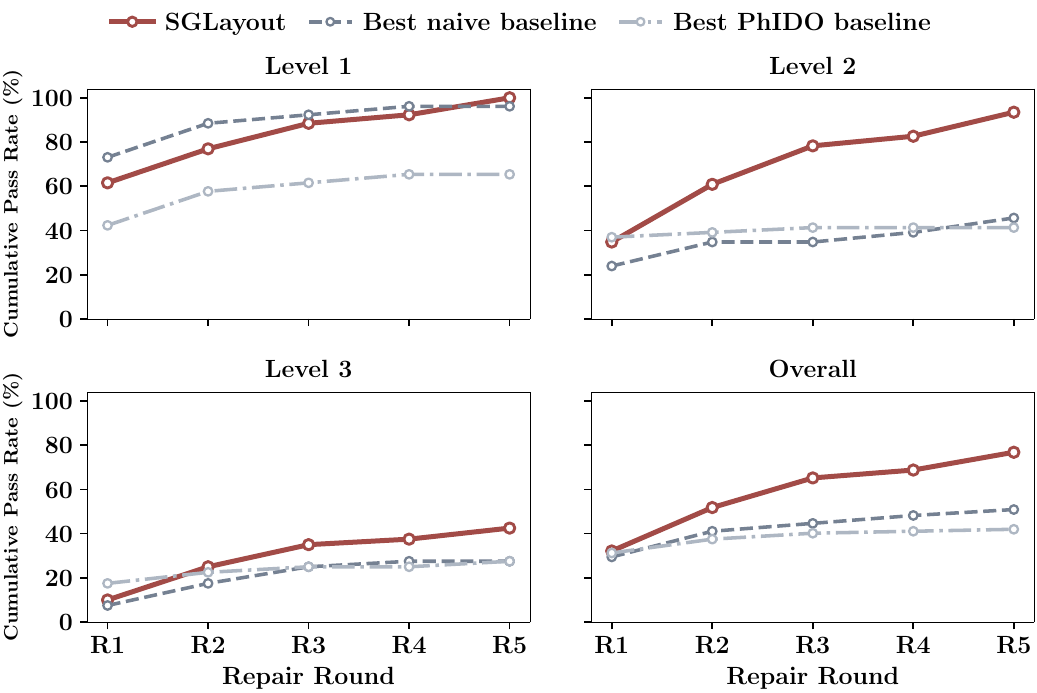}
\caption{Convergence by repair round for SGLayout, best naive baseline, and best PhIDO baseline across complexity levels.}
\label{fig:convergence-comparison}
\end{figure}

\Cref{tab:main-results} summarizes the main end-to-end results on the 112-task benchmark. Under the same evaluator stack and a matched five-trial budget, SGLayout achieves the best overall Final score of 76.79\%, outperforming the strongest naive baseline by 25.90 points and all PhIDO variants by an even larger margin. This result shows that SGLayout delivers the strongest overall convergence under a unified evaluation protocol. Compared with the naive baselines, the gain highlights the value of schema-guided repair, which grounds geometry-space feedback to executable code targets. Compared with PhIDO, the result further suggests that one-shot generation alone is insufficient for robust layout synthesis: although the best PhIDO variant remains competitive at Pass@1 on harder tasks, reaching 36.96\% on Level 2 and 17.50\% on Level 3, its Final scores rise only to 41.30\% and 27.50\% even after repeated independent attempts, indicating that iterative repair remains essential for reliably solving complex tasks.

The advantage of SGLayout grows with task difficulty. On Level 1, SGLayout reaches 100.00\% Final, only slightly above the strongest naive baseline (96.15\%). On harder tasks, however, the gap becomes substantial. On Level 2, SGLayout reaches 93.48\%, compared with 45.65\% for the best naive baseline and 41.30\% for the best PhIDO variant. On Level 3, SGLayout still achieves the best result at 42.50\%, while the best baseline remains at 27.50\%. These results indicate that SGLayout is most beneficial when failures involve stronger structural coupling and longer-range geometric dependencies, where successful refinement requires identifying root-cause code regions rather than addressing downstream symptoms.

Since one-shot generation by LLMs is inherently unstable and stochastic, our goal is to enable stable convergence through repair regardless of the quality of the initial generation. As further visualized in \Cref{fig:convergence-comparison}, although some baselines achieve comparable or even stronger Pass@1, SGLayout delivers larger gains from Pass@1 to Pass@3 and Final, especially on harder tasks. On Level 2, its performance improves from 34.78\% to 93.48\%, and on Level 3, from 10.00\% to 42.50\%, while the corresponding gains of the naive baselines and PhIDO are much smaller. This trend is consistent with the goal of SGLayout, which targets iterative repair rather than one-shot generation. Overall, the results support the central claim of this work: grounding geometry-space feedback in an execution-derived schema leads to more effective repair and more reliable end-to-end layout generation.

\subsection{Evaluator Reliability}
\label{sec:evaluator-reliability}
Because SGLayout relies on automatic evaluators to drive repair, their reliability must be validated independently. We therefore construct two dedicated reliability testbenches, one for the structural checker and one for the VLM evaluator, both using manually authored ground truth written and verified by experienced layout engineers. We report Accuracy, Precision, and Recall, treating the fail/violation class as the positive class: Accuracy measures overall agreement with ground truth, Precision measures the fraction of predicted failures that are true failures, and Recall measures the fraction of true failures that are detected. The detailed evaluation results are summarized in \Cref{tab:evaluator-reliability}.

For the structural intent checker, we build a 100-sample testbench from 20 representative queries spanning various task types. Each query includes one gold-pass sample and four controlled gold-fail variants covering practical structural error modes. This setup evaluates the checker in the same schema-driven environment as the main loop and verifies whether it can reliably isolate specific structural violations beyond coarse pass/fail labels. Under this suite, the checker achieves 94.00\% accuracy, 95.12\% precision, and 97.50\% recall, supporting its role as the primary intent gate.

For the VLM evaluator, we build a 56-sample visual testbench using 14 representative queries from various visually salient task types. Each query is rendered into four variants: gold-pass, fail-visual-a, fail-visual-b, and out-of-scope. The first three variants test clean-pass rejection and in-scope visual-failure recall, while the fourth keeps the overall visual appearance unchanged and introduces violations only in numeric or hierarchy-level properties that are not meant to be assessed by the VLM. In other words, this benchmark measures not only visual detection ability but also scope discipline. We use the exact same production stack to evaluate: \texttt{Qwen-VL-Max} with identical prompts and the dual raw and
annotated layout renderings. Under this setup, the VLM achieves 98.21\% accuracy, 100.00\% precision, and 96.43\% recall on in-scope visual failures, while showing no overreach on out-of-scope controls.

For the DRC evaluator, the rule deck is derived from the GDSFactory Generic PDK and focuses on foundational manufacturability constraints such as width, spacing, and enclosure. Global constraints, such as pattern density and grid alignment, are excluded as they are typically resolved during the chip-level stages. This scope strictly targets geometric violations rooted in code-level parameterization.

\begin{table}[t]
\centering
\caption{Reliability statistics of the Structural Checker and VLM Evaluator.}
\label{tab:evaluator-reliability}
\footnotesize
\begin{tabular}{l c c c c c}
\toprule
Evaluator & \# Samples & Accuracy (\%) & Precision (\%) & Recall (\%) \\
\midrule
Structural Checker & 100 & 94.00 & 95.12 & 97.50 \\
VLM Evaluator & 56 & 98.21 & 100.00 & 96.43 \\
\bottomrule
\end{tabular}
\end{table}

\subsection{Ablation Study}
We ablate the LLM-based feedback orchestrator to isolate its contribution to iterative repair. In this variant, only the LLM-based feedback orchestrator in the orchestration layer is removed, so the ablated variant can no longer perform cross-iteration orchestration or reason about the relation between the targeted code snippet and the complete program. A separate ablation of the schema-guided mechanism is not included because it is tightly coupled with the orchestration layer, and removing it would make the resulting pipeline largely equivalent to the naive baseline.

\begin{figure}[t]
\centering
\includegraphics[width=\columnwidth]{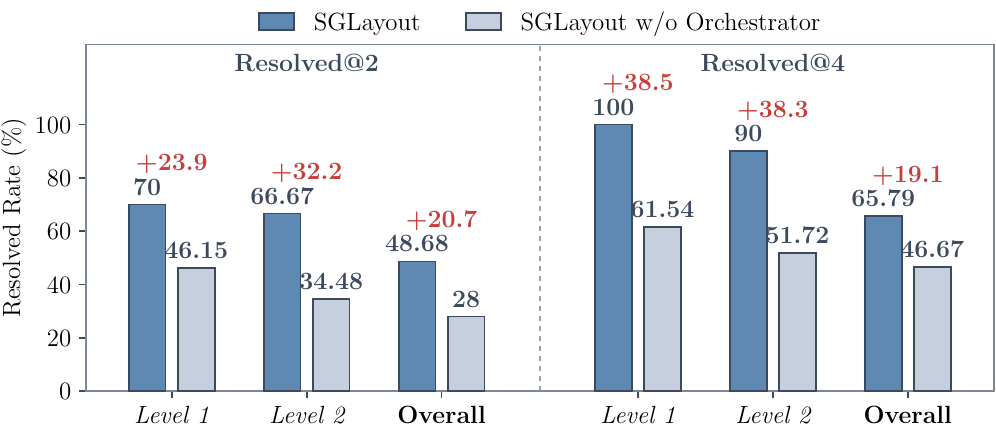}
\caption{Ablation study of the LLM-based feedback orchestrator on samples that fail in the initial generation round.}
\label{fig:ablation-results}
\end{figure}

Since the orchestrator is only activated when the initial generation fails, we evaluate both variants on the subset of first-round-failed samples. Here, Resolved@$k$ denotes the fraction of such samples that reach full pass status within $k$ subsequent refinement rounds. \Cref{fig:ablation-results} shows that the full system consistently outperforms the ablated variant, improving Resolved@2 from 28.00\% to 48.68\%, Resolved@4 from 46.67\% to 65.79\%, and reducing the average number of additional repair rounds from 2.20 to 2.00. The effect is even more pronounced on Level 2, where the full system reaches 66.67\% Resolved@2 and 90.00\% Resolved@4, compared with 34.48\% and 51.72\% for the ablated variant. These results indicate that SGLayout’s repair gains stem not only from targeted localization, but also from the feedback orchestrator’s ability to integrate cross-iteration state, reason about the logical role of targeted code snippets within the complete program, and derive actionable repair plans.

\subsection{Case Study \& Discussion}
\begin{figure}[t]
\centering
\begin{minipage}[t]{0.48\columnwidth}
\centering
\subfloat[Initial Layout\label{fig:case-study-a}]{
\includegraphics[width=1.0\linewidth,trim=18 14 18 14,clip]{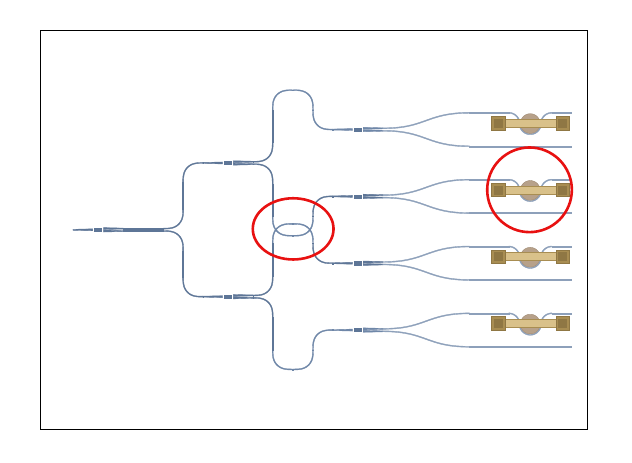}
}
\end{minipage}
\hspace{0.01\columnwidth}
\begin{minipage}[t]{0.48\columnwidth}
\centering
\subfloat[Intermediate Refinement\label{fig:case-study-b}]{
\includegraphics[width=1.0\linewidth,trim=18 14 18 14,clip]{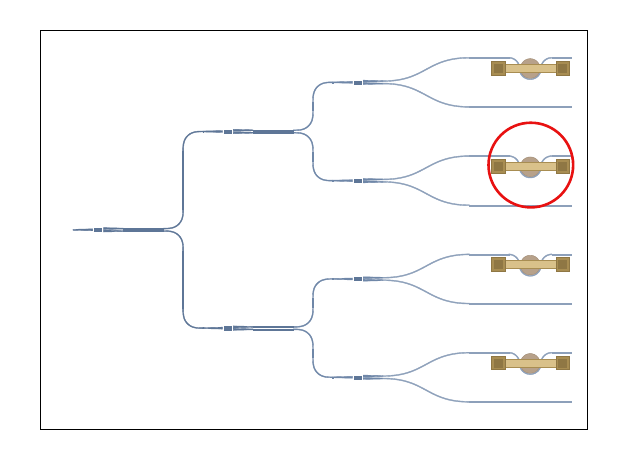}
}
\end{minipage}\\[0.25em]
\begin{minipage}[t]{0.48\columnwidth}
\centering
\subfloat[Naive Refinement Result\label{fig:case-study-c}]{
\includegraphics[width=1.0\linewidth,trim=19 22 19 22,clip]{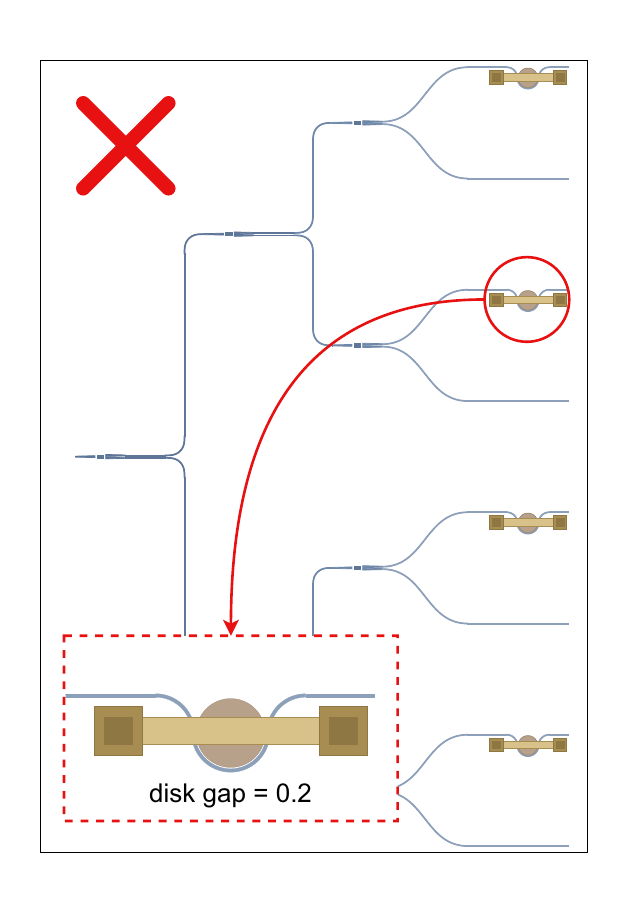}
}
\end{minipage}
\hspace{0.01\columnwidth}
\begin{minipage}[t]{0.48\columnwidth}
\centering
\subfloat[SGLayout Refinement Result\label{fig:case-study-d}]{
\includegraphics[width=1.0\linewidth,trim=19 22 19 22,clip]{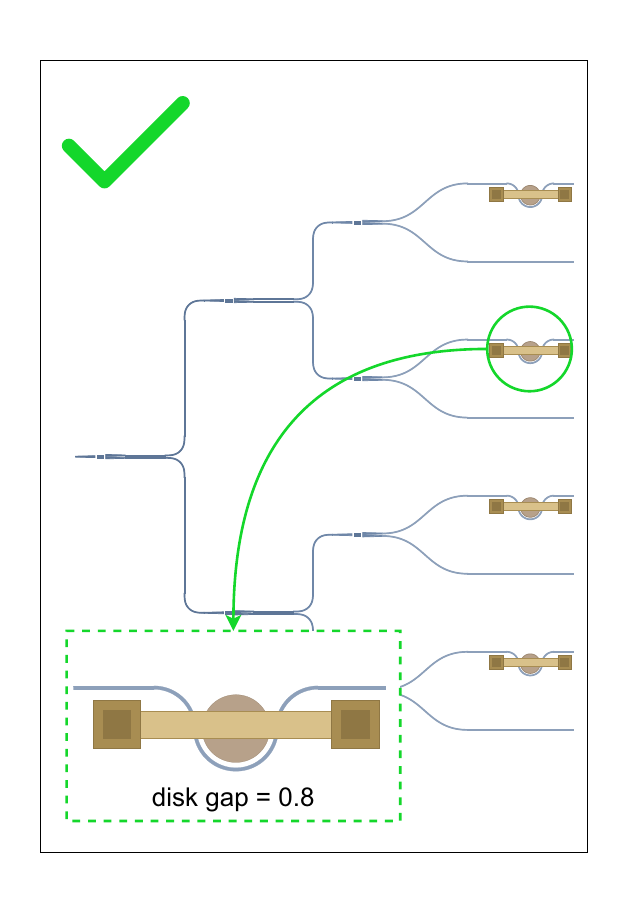}
}
\end{minipage}
\caption{Case study of iterative refinement on an 8-way splitter tree layout.}
\label{fig:case-study}
\end{figure}

\Cref{fig:case-study} shows a representative repair trajectory on an 8-way splitter tree layout with interleaved active and passive branches. The system first performs one-shot generation to produce an initial layout, in which the splitter tree is vertically under-expanded and the downstream bank pitch is too tight, leading to early failures in structural checking, as illustrated in \Cref{fig:case-study-a}.

Starting from the shared initial layout, both SGLayout and the naive baseline first address the same coarse structural bottleneck by increasing the splitter-tree vertical spacing, as shown in \Cref{fig:case-study-b}. The remaining failures, however, no longer arise from insufficient global branch separation, but from a repeated local spacing violation inside each active disk-heater branch: with \texttt{disk\_gap}=0.20, the disk-heater geometry and its adjacent waveguide remain too close under the minimum-spacing rule.

The two methods begin to diverge once progress on the DRC violations stalls. The naive baseline, shown in \Cref{fig:case-study-c}, keeps increasing the global parameter \texttt{spacing\_y} without changing the true error source. As a result, it remains trapped in the same repeated DRC violations and still fails after five rounds. In contrast, as shown in \Cref{fig:case-study-d}, SGLayout shifts the repair scope to the disk-heater branches after observing that repeated global spacing edits no longer resolve the DRC violations, and retunes \texttt{disk\_gap} from 0.20 to 0.80 while preserving the sparse-bank topology. This targeted repair allows the design to converge in Round~4. This example highlights the value of grounded repair when raw feedback does not directly expose the responsible code region.

More broadly, the current study focuses on executable text-to-layout generation rather than the full-custom physical design flow. Although the current instantiation is built on GDSFactory, the overall methodology exhibits strong transferability and is not tied to a single programming environment, PDK, or rule deck. The key abstractions of SGLayout are defined at the level of the generation–execution–evaluation–repair closed-loop itself rather than at the level of any specific API family, allowing the automatic workflow to transfer across parameterized layout frameworks and PDK settings as long as they expose comparable execution and evaluation signals. In this sense, SGLayout should be viewed as a general methodology for addressing the lack of spatial awareness in LLM-driven parameterized layout generation.

\section{Conclusion}
\label{sec:conclusion}
In this work, we presented SGLayout, a schema-guided closed-loop refinement framework for LLM-driven parameterized layout generation. 
To address the limited spatial awareness of LLMs, SGLayout introduces an execution-derived schema as the core alignment layer between code space and geometry space, allowing heterogeneous evaluator signals from structural checking, vision-language assessment, and DRC to be back-projected onto precise code-level repair targets, which are then coordinated by the orchestration layer into actionable repair plans, thereby enabling surgical closed-loop iterative refinement.
Results on a curated benchmark of 112 parameterized layout generation tasks show that SGLayout consistently surpasses strong baselines in generating layouts that satisfy both design intent and physical constraints.

\clearpage
{
\bibliographystyle{IEEEtran}
\bibliography{ref/Top,ref/reference}
}

\end{document}